\documentclass[twocolumn,superscriptaddress,aps,pra]{revtex4-2}

\makeatletter
\newcommand*{\rom}[1]{\expandafter\@slowromancap\romannumeral #1@}
\makeatother

\usepackage{color}
\usepackage{graphicx}
\usepackage{hyperref}
\usepackage{amsmath,amssymb}
\usepackage{epstopdf}
\usepackage{subcaption}
\usepackage{float}

\graphicspath{{figs/}}

\def\beq{\begin{equation}}
\def\eeq{\end{equation}}
\def\bea{\begin{eqnarray}}
\def\eea{\end{eqnarray}}

\begin{document}
\pagestyle{plain}

\title{Haldane-Inspired Generalized Statistics}

\author{M. H. Naghizadeh Ardabili}
\affiliation{Department of Physics, University of Mohaghegh Ardabili, P.O. Box 179, Ardabil, Iran}
\author{Omid Yahyayi Monem}
\affiliation{Department of Physics, University of Mohaghegh Ardabili, P.O. Box 179, Ardabil, Iran}
\author{Morteza Nattagh Najafi}
\affiliation{Department of Physics, University of Mohaghegh Ardabili, P.O. Box 179, Ardabil, Iran}
\author{Hosein Mohammadzadeh}
\email{mohammadzadeh@uma.ac.ir}
\affiliation{Department of Physics, University of Mohaghegh Ardabili, P.O. Box 179, Ardabil, Iran}

\begin{abstract}
We propose and study a generalized quantum statistical framework—referred to as alpha statistics—that continuously interpolates between Bose--Einstein and Fermi--Dirac statistics and naturally extends into the hyperbosonic regime for \(\alpha < 0\). Inspired by Haldane's exclusion statistics, this formulation introduces a modified occupation weight function that encodes effective statistical interactions via the parameter \(\alpha\). Using thermodynamic geometry, we analyze the sign and singular behavior of the thermodynamic curvature as a diagnostic of underlying interactions and phase structures. A crossover temperature \(T^*\), at which the curvature changes sign, marks the transition between effectively attractive (Bose-like) and repulsive (Fermi-like) statistical regimes. When expressed relative to the Bose--Einstein condensation temperature \(T_c\), the ratio \(T^*/T_c\) depends universally on \(\alpha\). For negative \(\alpha\), corresponding to hyperbosonic statistics, we find curvature singularities at specific fugacities, indicating modified condensation phenomena distinct from conventional Bose condensation. These results highlight the geometric and thermodynamic consequences of alpha statistics and establish a link between fractional exclusion principles and curvature-induced interaction signatures in statistical thermodynamics.
\end{abstract}

\maketitle

\section{Introduction}\label{1}

The foundational distributions of statistical mechanics—namely the classical Maxwell–Boltzmann (MB) and the quantum Bose–Einstein (BE) and Fermi–Dirac (FD) distributions—are instrumental in describing the equilibrium properties of many-body systems, spanning domains from condensed matter physics to cosmology. These distributions are inherently linked to the fundamental quantum statistics obeyed by particles: bosons, which can occupy the same quantum state without restriction, and fermions, which are constrained by the Pauli exclusion principle to single occupancy per state. This dichotomy is reflected in the symmetry properties of their respective wavefunctions—symmetric for bosons and antisymmetric for fermions—and in the commutation or anticommutation relations of their creation and annihilation operators \cite{pathria2011statistical,niven2005exact,sanchez2024reconstruction,chaichian2011fermibose}.

However, this strict classification into bosons and fermions is not exhaustive. In low-dimensional systems—particularly in two dimensions—particles may exhibit statistics interpolating between bosonic and fermionic behavior. These particles, termed anyons, have been theoretically predicted and experimentally observed in phenomena such as the fractional quantum Hall effect, where excitations carry fractional charge and obey fractional statistics \cite{wilczek1982quantum,tsui1982two,khare2005fractional,wilczek1990fractional,keilmann2011statistically}.

To formalize such intermediate statistics, Haldane introduced the concept of fractional exclusion statistics (FES), which generalizes the Pauli principle by quantifying how the available Hilbert space for a species changes with the addition of other particles \cite{haldane1991fractional}. In this framework, the statistical interaction is characterized by a parameter \(g\), where \(g=0\) corresponds to bosons and \(g=1\) to fermions, allowing continuous interpolation between the two extremes. The thermodynamics of systems obeying FES have been extensively studied, revealing behavior distinct from traditional quantum statistics \cite{haldane1991fractional,wu1994statistical,murthy1994thermodynamics,wu1995statistical,fukui1995haldane,byczuk1995universality}.

Beyond FES, other generalizations of quantum statistics have been proposed to describe systems with non-standard statistical behavior. These include \(q\)-deformed statistics—arising from quantum groups and non-extensive thermodynamics—and Kaniadakis statistics, which modify entropy and distribution functions to account for relativistic effects, with applications in astrophysics and econophysics \cite{kaniadakis2001non,tsallis1988possible,biedenharn1989q,kaniadakis2002power_law,kaniadakis2006relativistic}.

Despite these advancements, alternative frameworks are still needed to capture the nuances of systems exhibiting intermediate or unconventional statistics. In this work, we propose a novel generalized weight function that extends the conventional BE and FD distributions. Our approach introduces a parameter-dependent function that smoothly interpolates between the bosonic and fermionic limits, offering a unified framework to describe particles with intermediate statistics. We derive the corresponding distribution functions, analyze their thermodynamic properties, and compare our findings with existing models of generalized statistics. This new formulation serves as a versatile tool for exploring systems where conventional quantum statistics prove insufficient.

Thermodynamic geometry has emerged as a powerful tool for analyzing the underlying structure of thermodynamic systems, particularly those involving complex or interacting particle ensembles. This approach, pioneered by Weinhold and Ruppeiner, defines a Riemannian geometric structure on the thermodynamic state space, where the curvature of the associated metric encodes information about microscopic interactions and system stability.

In this formalism, the thermodynamic curvature—a scalar derived from the Riemann curvature tensor—plays a central interpretative role. Positive curvature is typically associated with attractive interactions (as in bosonic systems), while negative curvature suggests repulsive interactions (as in fermionic systems). A vanishing curvature indicates an ideal, non-interacting gas. Moreover, divergences in the curvature often signal phase transitions, corresponding to non-analytic behavior in thermodynamic quantities \cite{ruppeiner1979thermodynamics,ruppeiner1995riemannian,ruppeiner2010thermodynamic,ruppeiner2012thermodynamic}.

This geometric framework has been successfully applied to a wide array of systems obeying classical, quantum, and generalized statistical distributions—including Tsallis, Kaniadakis, Haldane–Wu, deformed, dual, and unified statistics \cite{ruppeiner1995riemannian,janyszek1990riemannian,mohammadzadeh2016perturbative,mehri2020thermodynamic,mirza2010thermodynamic,mirza2011thermodynamic,ebadi2020thermodynamic,esmaili2024thermodynamic}. These studies confirm that thermodynamic geometry effectively captures statistical interactions embedded in distribution functions, providing a complementary perspective to standard thermodynamic analysis.

In this paper, we extend this geometric analysis to the class of generalized statistics introduced in the following sections. By constructing the appropriate thermodynamic metric—typically defined in terms of the Hessian of entropy or free energy—we evaluate the curvature associated with our proposed distribution. This enables us to investigate intrinsic statistical interactions and potential critical behavior. We compare our findings with results from both classical and generalized statistics, highlighting distinctive features and advantages of our formulation.

The structure of the paper is as follows:  
Section~\ref{2} briefly reviews fractional exclusion (Haldane) statistics.  
Section~\ref{3} introduces a proposed weigh function that gives rise to a distribution function reminiscent of Haldane statistics.  
Section~\ref{4} explores the correspondence between Haldane statistics and the newly introduced alpha statistics.  
Section~\ref{5} discusses the relevant thermodynamic quantities for this formulation.  
Section~\ref{6} presents the associated thermodynamic geometry.  
Section~\ref{7} investigates the crossover temperature between bosonic-like and fermionic-like behavior—that is, the transition between attractive and repulsive statistical interactions.  
Section~\ref{8} provides a brief analysis of the case with negative \(\alpha\).  
Finally, Section~\ref{9} summarizes our conclusions.

\section{A brief review of Haldane statistics}\label{2}

In conventional quantum statistics, bosons are particles that can occupy the same quantum state without restriction on the number of particles per state, whereas fermions obey the Pauli exclusion principle, which dictates that each quantum state can be occupied by at most one particle. Haldane statistics generalize these two extremes by introducing a fractional exclusion parameter \( g \), which interpolates between bosonic and fermionic behavior. Specifically, when \( g = 0 \), the particles obey Bose--Einstein statistics, and when \( g = 1 \), they follow Fermi--Dirac statistics. Intermediate values \( 0 < g < 1 \) correspond to a class of particles obeying fractional exclusion statistics, allowing for the exploration of many-body systems with nontrivial correlations and emergent quantum phenomena \cite{haldane1991fractional}.

Consider a system where \( N_i \) indistinguishable particles occupy \( G_i \) distinct quantum states at energy level \( \varepsilon_i \). The number of possible microstates \( W_i \) for bosons and fermions is given respectively by \cite{pathria2011statistical}:
\begin{eqnarray}
W_{i}^B = \frac{(N_i + G_i - 1)!}{N_i! \, (G_i - 1)!},\\
W_{i}^F = \frac{G_i!}{(G_i - N_i)! \, N_i!}.
\end{eqnarray}

Haldane proposed an interpolating statistical framework that enables a continuous transition between Bose–Einstein and Fermi–Dirac statistics by introducing the concept of fractional exclusion statistics \cite{haldane1991fractional}. The proposed generalized statistical weight is:
\begin{equation}\label{wg}
W_{i}(g) = \frac{\left[G _i+ (N_i - 1)(1 - g)\right]!}{N_i! \, \left[G_i - gN_i - (1 - g)\right]!}.
\end{equation}

In this framework, continuously varying the statistical parameter \( g \) between 0 and 1 due to the fractional exclusion property changes the statistical behavior of the particles. Specifically, Haldane extended the Pauli exclusion principle by introducing the concept of fractional exclusion statistics. For fermions, the addition of each particle reduces the number of accessible single-particle states for the remaining particles by one. In contrast, for bosons, the addition of particles does not reduce the number of accessible states.

Haldane's generalization allows for a more flexible relation: instead of associating the addition of a single particle with the exclusion of exactly one available state, one may add multiple particles—two, three, or more—while reducing the number of accessible states by only one unit. If the system contains only a single species of indistinguishable particles, the change in the dimension of the accessible single-particle Hilbert space due to the addition of particles can be described by the following relation:
\begin{equation}
\Delta d = -g \Delta N,
\end{equation}
where \( \Delta d \) denotes the variation in the dimension of the accessible Hilbert space resulting from the addition of \( \Delta N \) particles to the system. The parameter \( g \), previously introduced, is referred to as the fractional parameter.

Given the number of states according to relation \ref{wg} and using the definition of entropy in terms of the number of states as 
\begin{equation}
S = k_B \ln W(g),
\end{equation}
where \( W(g)=\prod_{i}W_{i}(g) \), the distribution function of this type of particles was derived by Y. S. Wu \cite{wu1995statistical}. By maximizing the entropy subject to the constraints of fixed internal energy and total particle number, and by defining the average occupancy number as  
\begin{equation}
\frac{N_i}{G_i} = n_i(g),
\end{equation}
it can be shown that the distribution function satisfies the following relation:
\begin{equation}
(1 - g n_i)^g \left[1 + (1 - g) n_i \right]^{1 - g} = n_i \, e^{\frac{\epsilon_i - \mu}{kT}},
\end{equation}
where \( \mu \) is the chemical potential.

By defining a new function \( \rho_{i}(\zeta) \), with \( \zeta \equiv \exp\left(\frac{\epsilon_i - \mu}{kT}\right) \), in terms of the distribution function as
\begin{equation}
n_i(g) = \frac{1}{\rho_{i}(\zeta) + g},
\end{equation}
it can be shown that this function satisfies the following implicit relation:
\begin{equation}
\rho_{i}(\zeta)^g \left[1 + \rho_{i}(\zeta) \right]^{1 - g} = \zeta.
\end{equation}

The fugacity is defined as \( z = e^{\mu / kT} \), which implies \( \zeta = z^{-1} \exp(\beta \epsilon_i) \), where \( \beta = 1/kT \). Therefore, the distribution function for a given value of \( g \) can be expressed as \cite{haldane1991fractional,wu1995statistical}
\begin{equation}
n_i(g) = f(g, \beta, z, \epsilon_i), 
\end{equation}
which emphasizes that the distribution function depends on the fractional parameter, temperature, fugacity, and the energy of the states.

\section{Haldane-Inspired Statistics}\label{3}

In this section, we aim to generalize the enumeration of possible configurations for distributing \( N_i \) particles among \( G_i \) available states, employing an approach that is fundamentally different from that of Haldane. This generalized formulation introduces a fractional parameter \( 0 \le \alpha \le 1 \), where the limiting cases \( \alpha = 0 \) and \( \alpha = 1 \) correspond to conventional bosonic and fermionic statistics, respectively. For intermediate values of \( \alpha \), the model gives rise to a novel form of fractional statistics. The number of accessible configurations is proposed to follow the relation:
\begin{equation}\label{Walpha}
W_{i}(\alpha) = W_{i}^{B} \left( \frac{W_{i}^{F}}{W_{i}^{B}} \right)^{\alpha},
\end{equation}
where \( W_i^{(B)} \) and \( W_i^{(F)} \) represent the number of configurations in the bosonic and fermionic limits, respectively.

The ratio of the two configuration numbers in Eq.~\eqref{Walpha} serves as a key quantity for interpolating between the bosonic and fermionic cases. In the limiting values of \( \alpha \), this expression naturally reproduces the conventional results. More precisely, the ratio is given by:
\begin{equation}
\frac{W_{i}^{F}}{W_{i}^{B}} = \frac{G_i \left[(G_i - 1)!\right]^2}{(G_i - N_i)! (N_i + G_i - 1)!}.
\end{equation}

Thus, Eq.~\eqref{Walpha} can equivalently be written in the form:
\begin{equation}
W_{i}(\alpha) = \left(W_i^{B}\right)^{1 - \alpha} \left(W_i^{F}\right)^{\alpha}.
\end{equation}

\begin{figure}[H]
    \centering
    \includegraphics[width=8.5cm, height=8.5cm]{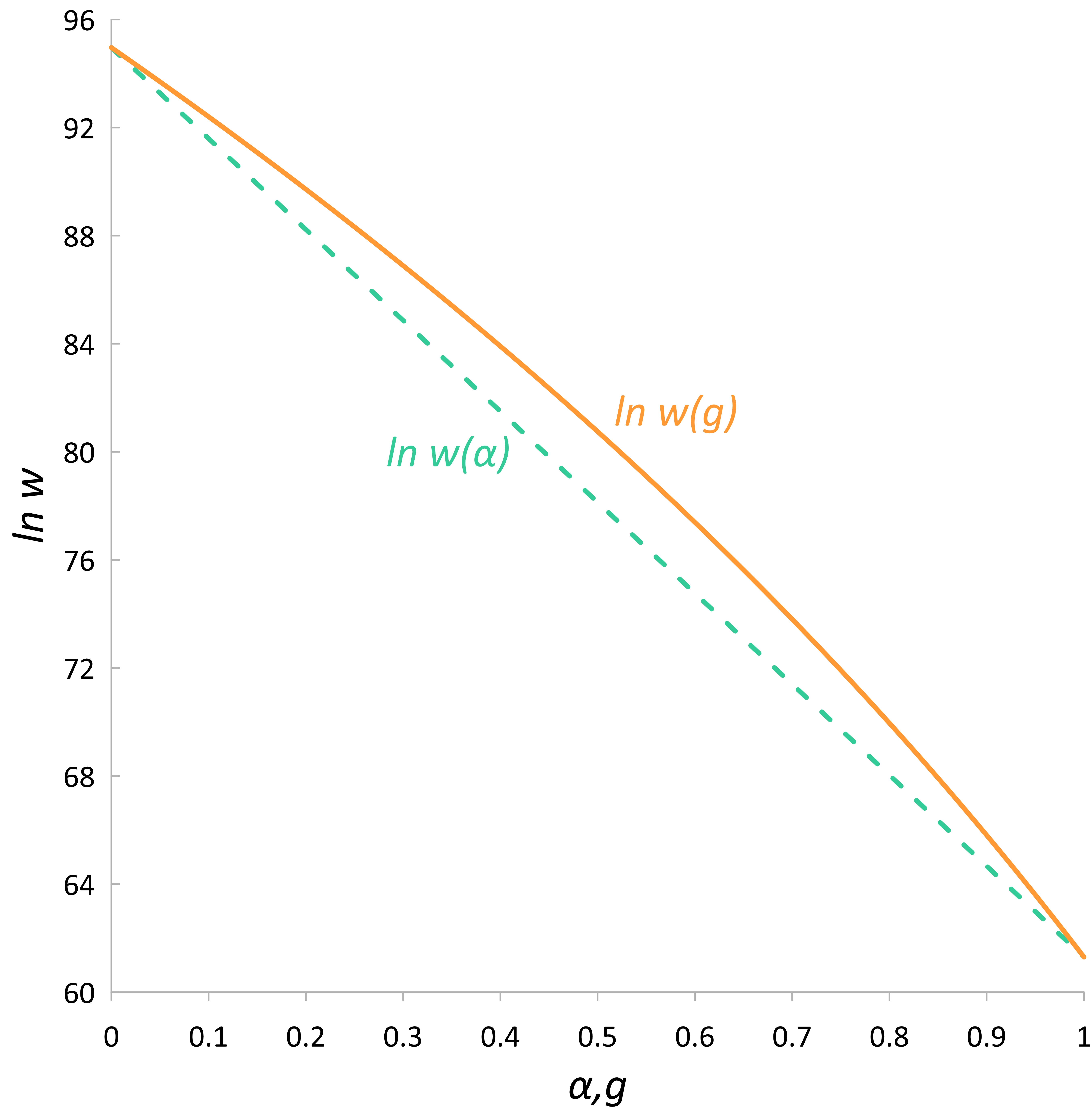}  
    \caption{\footnotesize The number of possible configurations for distributing \( N_i = 55 \) particles among \( G_i = 94 \) states as a function of the fractional parameter \(\alpha\). The solid line represents the Haldane generalization, and the dashed line represents the proposed generalization.}
    \label{fig1}
\end{figure}

In Fig. (\ref{fig1}), the number of configurations based on Haldane's statistics and those based on the proposed model are plotted as functions of the generalized fractional parameter \( \alpha \), for a fixed number of particles \( N_i = 55 \) and available states \( G_i = 94 \).

The weight function introduced in this model leads to a generalized entropy that is a linear combination of the bosonic and fermionic entropies. The total weight function is \( W(\alpha)=\prod_{i}W_{i}(\alpha) \) and therefore, the total entropy is given by:
\begin{equation}\label{wabf}
S(\alpha) = k_B \ln W(\alpha) = (1 - \alpha) S_B + \alpha S_F,
\end{equation}
where \( S_B \) and \( S_F \) denote the entropy associated with bosonic and fermionic statistics, respectively.

We now examine the distribution function corresponding to the proposed statistical weight in the thermodynamic limit.

\begin{widetext}
Applying Stirling’s approximation to Eq.~\eqref{wabf}, the logarithm of the total number of configurations becomes:
\begin{align}
\ln W(\alpha) =\;& (1 - \alpha)\sum_{i} \left[ (N_i + G_i - 1) \ln(N_i + G_i - 1) 
                 - (G_i - 1) \ln(G_i - 1) \right] \notag \\
              &+ \alpha \sum_{i} \left[ -(G_i - N_i) \ln(G_i - N_i) \right] 
               - \sum_{i} N_i \ln N_i.
\end{align}
\end{widetext}

Applying the method of Lagrange multipliers and imposing the constraints of fixed total particle number and internal energy, we maximize the entropy and obtain the distribution function. The average occupation number \( n_i \) satisfies:
\begin{equation}
(1 + n_i)^{1 - \alpha} (1 - n_i)^{\alpha} = n_i e^{\beta \epsilon_i - \gamma},
\end{equation}
where \( \beta = 1/(k_B T) \), \( \epsilon_i \) is the energy of state \( i \), and \( \gamma \) is the Lagrange multiplier related to particle number conservation.

For \( \alpha = 0 \) and \( \alpha = 1 \), this equation reduces to the Bose-Einstein and Fermi-Dirac distributions, respectively. For intermediate values of \( \alpha \), it describes a mixed (fractional) distribution. Defining a new function \( u_i \) such that:
\begin{equation}\label{nu}
n_{i}(\alpha) = \frac{1}{u_i + \alpha},
\end{equation}
we find that \( u_i \) satisfies the following transcendental equation:
\begin{equation}\label{ud}
(u_i + \alpha + 1)^{1 - \alpha} (u_i + \alpha - 1)^{\alpha} = z^{-1} e^{\beta \epsilon_i},
\end{equation}
where \( z = e^{\gamma} \) is the fugacity. Due to the dependence of the distribution function on the fractional parameter \( \alpha \), we will henceforth refer to it as the "alpha statistics".

Equations (11) and (7), which are derived from the alpha and Haldane statistics respectively, show notable similarities. In particular, for \( \alpha = 0 \) and \( g = 0 \), corresponding to the Bose-Einstein distribution, both yield the same result, \( \rho_{i} = u_i = z^{-1} \exp({\beta \epsilon_i})-1 \). Also, for \( \alpha = 1 \) and \( g = 1 \), we obtain \( \rho_{i} = u_i = z^{-1} \exp({\beta \epsilon_i}) \).

\begin{figure}[h]
    \centering
    \includegraphics[scale=0.39]{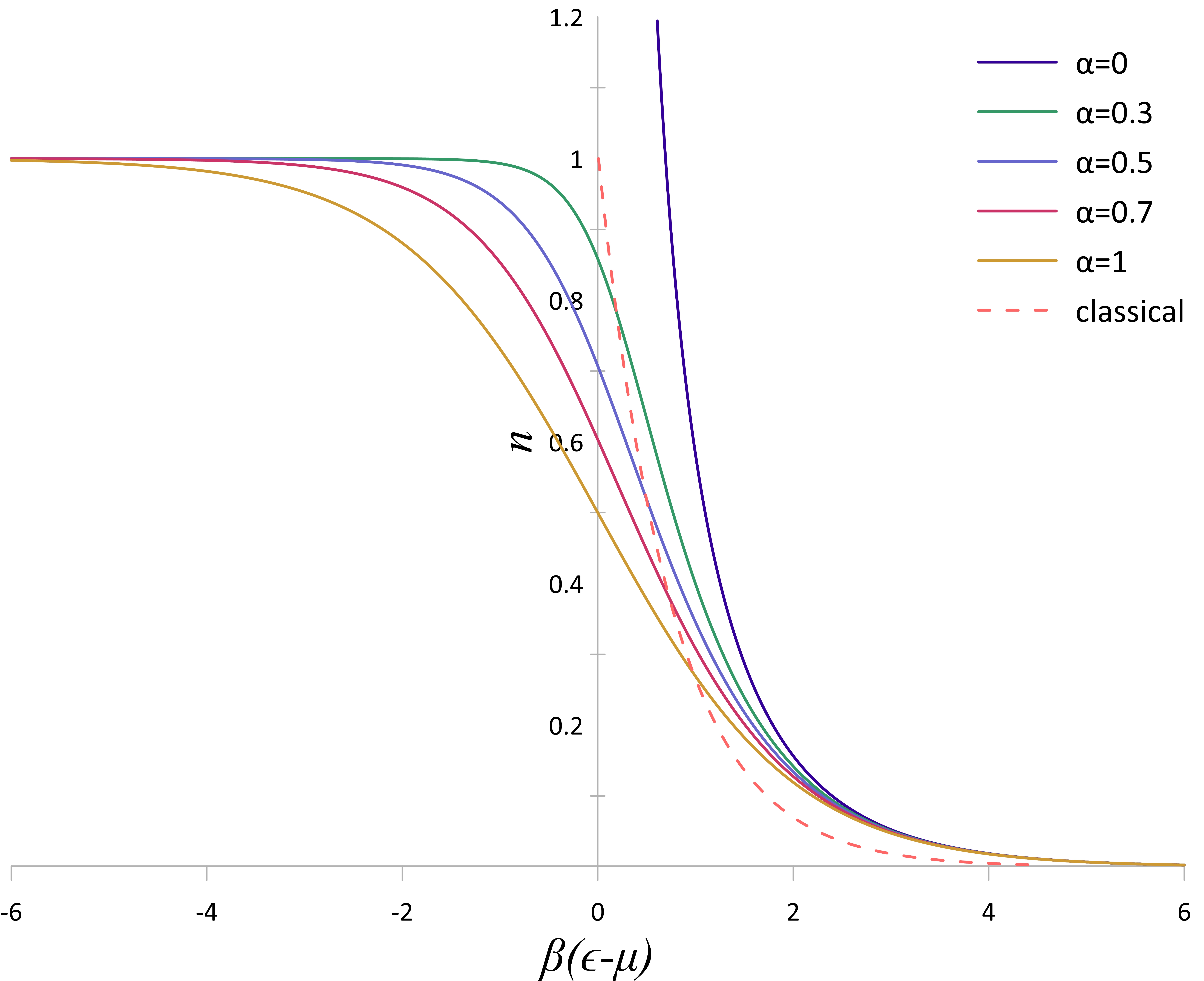}
    \caption{Distribution function of alpha statistics as a function of \( \beta(\epsilon - \mu) \) for different values of \( \alpha \).}
    \label{distribution}
\end{figure}

 \section{Correspondence Between Haldane and alpha Statistics}\label{4}
Suppose we apply the following variable transformation to the equation below.:
\begin{equation}
2v_i = u_i + \alpha - 1
\end{equation}
Rewriting Eq. (\ref{ud}) in terms of the new variable \( v_i \) leads to the following equation:
\begin{equation}
(1 + v_i)^{1 - \alpha} v_i^{\alpha} = (2z)^{-1} \exp(\beta \epsilon_i)
\end{equation}
This equation is specifically equivalent to equation (7) for the Haldane distribution, with the correspondence that Haldane's fractional parameter \( g \) is replaced by \( \alpha \) and the fugacity \( z \) is replaced by \( z' = 2z \). Now, the distribution function in terms of \( v_i \) can be written as follows:
\begin{equation}
n_{i}(\alpha ) = \frac{1}{2v_i + 1}
\end{equation}
We aim to establish a relationship between the Haldane distribution function and the proposed new distribution function. Given that 
\( v_i(\epsilon_i, \beta, z) = \rho_{i}(\epsilon_i, \beta, z^2) \), 
we can derive that:
\begin{equation}\label{ngalpha}
n_{i}(\alpha ,z, \beta, \epsilon_i) = \frac{n_{i}(g ,z^2, \beta, \epsilon_i)}{2 + (1 - g) n_{i}(g,z^2, \beta, \epsilon_i)} \Big|_{g=\alpha}
\end{equation}
Thus, by identifying the parameters of the two distributions, a correspondence between the distribution functions is established via relation (\ref{ngalpha}).
 \section{Thermodynamic quantities}\label{5}
In this section, using the evaluated distribution function given by Eqs.~(\ref{nu}) and (\ref{ud}), we derive the thermodynamic quantities, such as the internal energy and the total number of particles. We consider an ideal gas confined in a $D$-dimensional box, where the particles obey alpha statistics and follow the following energy-momentum dispersion relation:
\begin{equation}
\epsilon = a p^{\sigma},
\end{equation}
where \( p \) is the momentum of the particles, \( \sigma \) is the dispersion exponent, and \( a \) is a proportionality constant.

We examine two limiting cases: the non-relativistic and ultra-relativistic regimes. In the non-relativistic limit, the energy is given by \(\epsilon = {p^2}/{2m}\), where \( m \) is the mass of the particles. Here, the dispersion exponent is \( \sigma = 2 \) and \( a = {1}/{2m} \). In the ultra-relativistic limit, the energy is given by \(\epsilon = pc\), where \( c \) is the speed of light. In this case, the dispersion exponent is \( \sigma = 1 \) and \( a = c \).

In $D$ dimensions, the density of single-particle states, neglecting spin, is given by:
\begin{equation}
\Omega(\epsilon) = \frac{A^D}{\Gamma\left(\frac{D}{\sigma}\right)} \epsilon^{\frac{D}{\sigma} - 1},
\end{equation}
where \( A \) is a constant defined as:
\begin{equation}
A = \frac{L \sqrt{\pi}}{a^{\frac{1}{\sigma}} h},
\end{equation}
and \( L^D \) represents the volume of the $D$-dimensional box. For simplicity, we set \( A = 1 \) in the subsequent calculations.

\begin{widetext}
Replacing $w_{i}=u_{i}+\alpha$ in Eq.~(\ref{ud}) and omitting the subscript $i$ in the continuum limit, we obtain:
\begin{eqnarray}
   && \beta (\epsilon - \mu) = (1 - \alpha) \ln(w + 1) + \alpha \ln(w - 1),\label{weq} \\
   && \beta\, d\epsilon = (1 - \alpha) \frac{dw}{w + 1} + \alpha \frac{dw}{w - 1}.
\end{eqnarray}

By performing a change of variable from \( \epsilon \) to \( w \), the internal energy and total number of particles in the thermodynamic limit are given by:
\begin{eqnarray}
U &=& \int_0^\infty \epsilon\, n(\epsilon)\, \Omega(\epsilon) \, d\epsilon \nonumber\\
&=& \frac{A^D}{\Gamma\left( \frac{D}{\sigma} \right)} \beta^{-\left( \frac{D}{\sigma} + 1 \right)} 
\int_{w_0}^{\infty} \frac{1}{w} \left[ \ln\left( (w + 1)^{1-\alpha} (w - 1)^{\alpha} z \right) \right]^{\frac{D}{\sigma}} 
\frac{w + 2\alpha - 1}{w^2 - 1} \, dw, \label{ue}\\
N &=& \int_0^\infty n(\epsilon)\, \Omega(\epsilon) \, d\epsilon \nonumber\\
&=& \frac{A^D}{\Gamma\left( \frac{D}{\sigma} \right)} \beta^{-\frac{D}{\sigma}} 
\int_{w_0}^{\infty} \frac{1}{w} \left[ \ln\left( (w + 1)^{1-\alpha} (w - 1)^{\alpha} z \right) \right]^{\frac{D}{\sigma} - 1} 
\frac{w + 2\alpha - 1}{w^2 - 1} \, dw,\label{ne}
\end{eqnarray}

where \( w_0 \) corresponds to the value of \( w \) at \( \epsilon = 0 \), and is determined by:
\begin{equation}\label{w0}
(1 - \alpha) \ln(w_0 + 1) + \alpha \ln(w_0 - 1) = -\beta \mu.
\end{equation}

We note that an analytical solution to obtain the explicit form of the distribution function is generally not possible, except for specific values of \( \alpha \). Changing the variable \( \epsilon \) to \( w \) enables us to evaluate the integrals numerically.
Therefore, for a given value of \( \alpha \), temperature \( T = {1}/{k_B \beta} \), and fugacity \( z \), the parameter \( w_0 \) can be determined, allowing the internal energy \( U \) and particle number \( N \) to be computed accordingly.
\end{widetext}
 \section{Thermodynamic geometry}\label{6}
Ruppeiner and Weinhold introduced thermodynamic geometry as a powerful framework for analyzing the structure and behavior of thermodynamic systems from a geometric perspective~\cite{ruppeiner1979thermodynamics,weinhold1975metric}. In this approach, the space of thermodynamic parameters is treated as a Riemannian manifold, where geometric quantities such as curvature offer insights into the nature of microscopic interactions.

The Ruppeiner metric is defined as the negative Hessian of entropy with respect to extensive variables such as internal energy, volume, and particle number~\cite{ruppeiner1979thermodynamics}. In contrast, Weinhold proposed a metric based on the second derivatives of the internal energy with respect to these extensive variables~\cite{weinhold1975metric}. These two metrics are related by a conformal transformation involving the system temperature~\cite{salamon1984relation}.

Legendre transformations facilitate the transition between different thermodynamic potentials—such as the Helmholtz and Gibbs free energies—by exchanging the roles of extensive and intensive variables. In addition to these classical metrics, the Fisher–Rao information metric emerges naturally within the framework of information geometry. It is constructed from the second derivatives of the logarithm of the partition function with respect to non-extensive parameters, typically the inverse temperature \(\beta = 1 / k_B T\) and the dimensionless chemical potential parameter \(\gamma = -\mu / k_B T\)~\cite{ruppeiner1995riemannian,janyszek1990riemannian,brody1995geometrical,amari2000methods}:

\begin{equation}
g_{ij} = \partial_i \partial_j \ln \mathcal{Z},
\end{equation}

where \(\partial_i\) denotes partial differentiation with respect to the \(i\)-th non-extensive thermodynamic variable, and \(\mathcal{Z}\) is the partition function. For classical and quantum ideal gases, the logarithm of the partition function typically depends on \(\beta\) and \(\gamma\), with the system volume held fixed. Consequently, the geometry of thermodynamic fluctuations can often be effectively described in a two-dimensional parameter space.

The Christoffel symbols, which act as connection coefficients, are expressed in terms of the metric tensor components as \cite{schutz1980}:

\begin{equation}
\Gamma^i_{jk} = \frac{1}{2} g^{im} \left( \partial_k g_{mj} + \partial_j g_{mk} - \partial_m g_{jk} \right),
\end{equation}

where \(g^{im}\) denotes the components of the inverse metric tensor, and \(\partial_k g_{ij}\) represents the partial derivative of the metric component with respect to the parameter \(k\).

The Riemann curvature tensor, which encodes the intrinsic curvature of the thermodynamic manifold, is defined by:

\begin{equation}
R^i_{jkl} = \partial_k \Gamma^i_{lj} - \partial_l \Gamma^i_{kj} + \Gamma^i_{km} \Gamma^m_{lj} - \Gamma^i_{lm} \Gamma^m_{kj}.
\end{equation}

From the Riemann tensor, the Ricci tensor, a second-rank contraction, is obtained as:

\begin{equation}
R_{ij} = R^m_{imj},
\end{equation}

and the scalar curvature, often referred to as the thermodynamic scalar curvature, is given by:

\begin{equation}
R = g^{ij} R_{ij}.
\end{equation}

In the case of a two-dimensional thermodynamic parameter space, the Ricci scalar simplifies significantly.

The system under consideration involves two fluctuating thermodynamic parameters; therefore, the dimension of the thermodynamic parameter space is equal to two. The scalar curvature is given by the following relation \cite{ruppeiner1995riemannian}:
\begin{equation}
R = \frac{R_{1212}}{2\det g},
\end{equation}
where \( R_{1212} \) is a component of the Riemann curvature tensor, and \( \det g \) is the determinant of the metric tensor.

Janyszek and Mrugala demonstrated that when the metric components are expressed exclusively as second derivatives of a specific thermodynamic potential, the thermodynamic curvature can be expressed in terms of second and third derivatives. Furthermore, the sign convention used here is consistent with that adopted by Janyszek and Mrugala.

For a two-dimensional thermodynamic space spanned by the parameters \( \beta \) and \( \gamma \), the Ricci scalar can be written as:
\begin{widetext}
\begin{equation}
R = - \frac{
\begin{vmatrix}
g_{\beta\beta} & g_{\beta\gamma} & g_{\gamma\gamma} \\
 g_{\beta\beta,\beta} & g_{\beta\gamma,\beta} & g_{\gamma\gamma,\beta} \\
 g_{\beta\beta,\gamma} & g_{\beta\gamma,\gamma} &  g_{\gamma\gamma,\gamma}
\end{vmatrix}
}{2
\begin{vmatrix}
g_{\beta\beta} & g_{\beta\gamma} \\
g_{\beta\gamma} & g_{\gamma\gamma}
\end{vmatrix}^2}.
\end{equation}
In the context of thermodynamic geometry, particles that follow the classical Maxwell--Boltzmann distribution are associated with a flat thermodynamic space and zero curvature, indicating a neutral statistical interaction. For bosons, the thermodynamic curvature is positive, signifying an attractive statistical interaction, whereas for fermions, the curvature is negative, reflecting a repulsive statistical interaction. Furthermore, thermodynamic curvature serves as a useful tool for identifying phase transitions. For example, the condensation point of an ideal Bose gas, which occurs at the critical fugacity \( z = 1 \), is characterized by a singularity in the thermodynamic curvature \cite{may2013thermodynamic,ruppeiner2012thermodynamic,dey2012information}. 

The thermodynamic geometry of various generalized statistics has also been investigated, including fractional exclusion statistics, Polychronakos statistics, Gentile statistics, deformed statistics, Kaniadakis statistics, nonextensive statistics, quantum unified statistics, and Mittag-Leffler-based statistics.
 \cite{mirza2010thermodynamic,mirza2011condensation,oshima1999riemann,mirza2011thermodynamic,mohammadzadeh2017thermodynamic,mehri2020thermodynamic,mohammadzadeh2016perturbative,esmaili2024thermodynamic,seifi2025intrinsic}.  

In the following, we construct the thermodynamic parameter space and compute the thermodynamic curvature of alpha statistics. To begin, we calculate the metric elements using the equations for internal energy and total particle number, (\ref{ue}) and (\ref{ne}). The metric tensor elements are determined as follows:
\begin{eqnarray}
g_{\beta\beta} &=& \frac{\partial^2 \ln \mathcal{Z}}{\partial \beta^2} = -\left(\frac{\partial U}{\partial \beta}\right)_{\gamma} \nonumber \\
&&= \frac{ \frac{D}{\sigma} + 1 }{ \Gamma\left( \frac{D}{\sigma} \right) \cdot \beta^{\frac{D}{\sigma} + 2} }
\left( \int_{w_{0}}^{\infty} \frac{1}{w} 
\left( \ln \left( (w+1)^{1-\alpha} (w-1)^{\alpha} z \right) \right)^{\frac{D}{\sigma}} 
\cdot \left( \frac{w + 2\alpha - 1}{w^2 - 1} \right) dw \right) ,\label{gbb}\\
g_{\beta \gamma} &=& g_{\gamma \beta} = \frac{\partial^2 \ln \mathcal{Z}}{\partial \beta \partial \gamma} = -\left(\frac{\partial N}{\partial \beta}\right)_{\gamma} \nonumber \\
&&= \frac{\frac{D}{\sigma}}{\Gamma\left(\frac{D}{\sigma}\right) \cdot \beta^{\frac{D}{\sigma}+1}} 
\left( \int_{w_{0}}^{\infty} \frac{1}{w} 
\left( \ln \left( (w+1)^{1-\alpha} (w-1)^{\alpha} z \right) \right)^{\frac{D}{\sigma}-1} 
\cdot \left( \frac{w + 2\alpha - 1}{w^2 - 1} \right) dw \right), \label{gbg}\\
g_{\gamma \gamma} &=& \frac{\partial^2 \ln \mathcal{Z}}{\partial \gamma^2} = -\left(\frac{\partial N}{\partial \gamma}\right)_{\beta} \nonumber \\
&&= \frac{1}{\Gamma\left(\frac{D}{\sigma}\right)  \beta^{\frac{D}{\sigma}}} 
\left( \int_{w_{0}}^{\infty} \frac{1}{w^2} 
\left( \ln \left( (w+1)^{1-\alpha} (w-1)^{\alpha} z \right) \right)^{\frac{D}{\sigma}-1} dw \right),\label{ggg}
\end{eqnarray}
where \( w_0 \) is obtained using the relation~(\ref{w0}), and the last metric element, \( g_{\gamma\gamma} \), is calculated using the particle number equation~(\ref{ne}) and the chain rule of differentiation as follows:
\begin{equation}\label{cr}
\left( \frac{\partial N}{\partial \gamma} \right)_{\beta} = \left(\frac{\partial N}{\partial w}\right)_{\beta} \left(\frac{\partial w}{\partial \gamma}\right)_{\beta},
\end{equation}
\[
\hspace{-10cm} 
\text{where, } ({\partial w}/{\partial\gamma})_{\beta} \text{ is calculated using the relation  \ref{weq}} .
\] 
Differentiating Eqs.~(\ref{gbb}), (\ref{gbg}), and (\ref{ggg}) with respect to \( \beta \) and \( \gamma \), we obtain the required terms to evaluate the thermodynamic curvature as follows:
\begin{eqnarray}
g_{\beta\beta,\beta} &=& \frac{-\left(\frac{D}{\sigma} + 1\right) \left(\frac{D}{\sigma} + 2\right)}{\Gamma\left(\frac{D}{\sigma}\right) \beta^{\frac{D}{\sigma} + 3}} 
 \int_{w_0}^{\infty} \frac{1}{w} 
\left( \ln \left( (w+1)^{(1-\alpha)} (w-1)^{\alpha} z \right) \right)^{\frac{D}{\sigma}} 
 \left( \frac{w+2\alpha-1}{w^2-1} \right) dw, \\
g_{\beta\beta,\gamma} &=& g_{\beta\gamma,\beta} = g_{\gamma\beta,\beta} = \frac{-\frac{D}{\sigma} \left(\frac{D}{\sigma} + 1\right)}{\Gamma\left(\frac{D}{\sigma}\right) \beta^{\frac{D}{\sigma} + 2}} 
\int_{w_0}^{\infty} \frac{1}{w} 
\left( \ln \left( (w+1)^{(1-\alpha)} (w-1)^{\alpha} z \right) \right)^{\frac{D}{\sigma} - 1} 
\left( \frac{w+2\alpha-1}{w^2-1} \right) dw, \\
g_{\beta\gamma,\gamma} &=& g_{\gamma\beta,\gamma} = g_{\gamma\gamma,\beta} = \frac{-\frac{D}{\sigma}}{\Gamma\left(\frac{D}{\sigma}\right) \beta^{\frac{D}{\sigma} + 1}} 
 \int_{w_0}^{\infty} \frac{1}{w^2} 
\left( \ln \left( (w+1)^{(1-\alpha)} (w-1)^{\alpha} z \right) \right)^{\frac{D}{\sigma} - 1} dw, \\
g_{\gamma\gamma,\gamma} &=& \frac{-1}{\Gamma\left(\frac{D}{\sigma}\right) \beta^{\frac{D}{\sigma}}} 
\int_{w_0}^{\infty}
\frac{w^5 - 4w^3 + (-4\alpha + 2)w^2 + 3w + 4\alpha - 2}{(2\alpha + w - 1)^2 w^3 (w^2 - 1)} 
\left( \ln \left( (w+1)^{(1-\alpha)} (w-1)^{\alpha} z \right) \right)^{\frac{D}{\sigma} - 1}
 dw.
\end{eqnarray}

\end{widetext}
\( G_{\gamma\gamma,\gamma} \) is obtained by differentiating \( G_{\gamma\gamma} \) with respect to \(\gamma\), using the chain rule as described similarly in Eq.~(\ref{cr}).

\begin{figure}[h]
    \centering
    \includegraphics[scale=0.37]{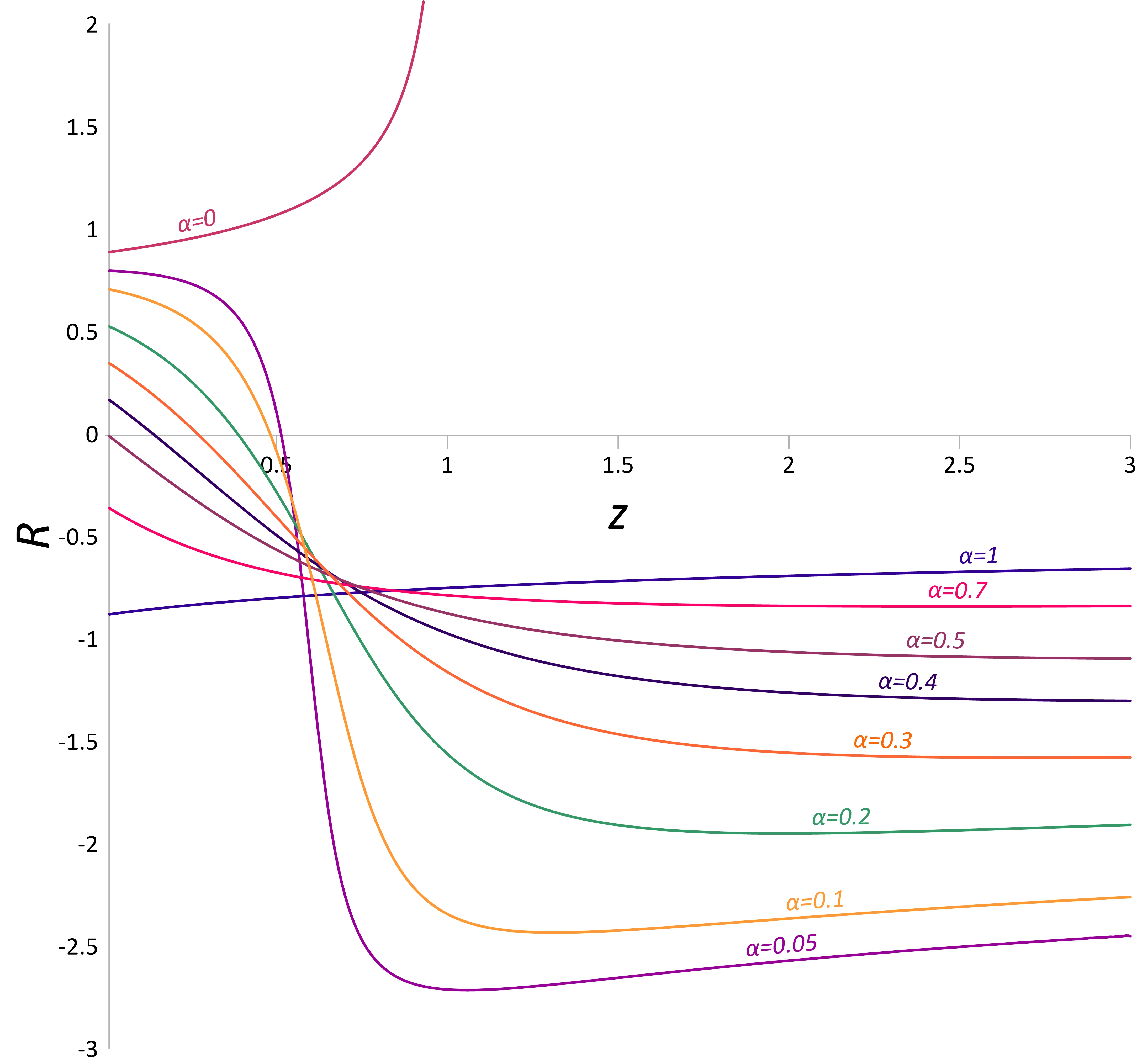}
    \caption{Thermodynamic curvature as a function of fugacity for isothermal processes (\(\beta = 1\)) in a 3-dimensional ideal \(\alpha\)-statistics gas with \(\sigma = 2\) (non-relativistic regime).}
    \label{fig2}
\end{figure}

Using the derived relations, all the necessary expressions are now available to determine the thermodynamic curvature and analyze the system's behavior in different dimensions for both non-relativistic and ultra-relativistic particles. Figure~\ref{fig2} shows the thermodynamic curvature of an ideal gas obeying alpha statistics as a function of fugacity for \(D/\sigma = 3/2\) under isothermal conditions. The behavior differs notably between the intervals \(0 < \alpha < 0.5\) and \(0.5 \le \alpha \le 1\). Specifically, for \(\alpha = 0\), the curvature corresponds to that of an ideal Bose gas, while for \(\alpha = 1\) it corresponds to that of a Fermi gas. For \(0.5 \le \alpha \le 1\), the thermodynamic curvature remains negative across the entire physical range, indicating that the intrinsic statistical interactions are repulsive and that the system exhibits fermion-like behavior. Furthermore, analysis of Figure~\ref{fig2} reveals that for \(0 < \alpha < 0.5\) there exists a specific fugacity \(z = z^*\) at which the thermodynamic curvature changes sign. This implies that the statistical interaction is repulsive (negative thermodynamic curvature) for \(z < z^*\) and becomes attractive (positive thermodynamic curvature) for \(z > z^*\).

\begin{figure}[h]
    \centering
    \includegraphics[scale=0.37]{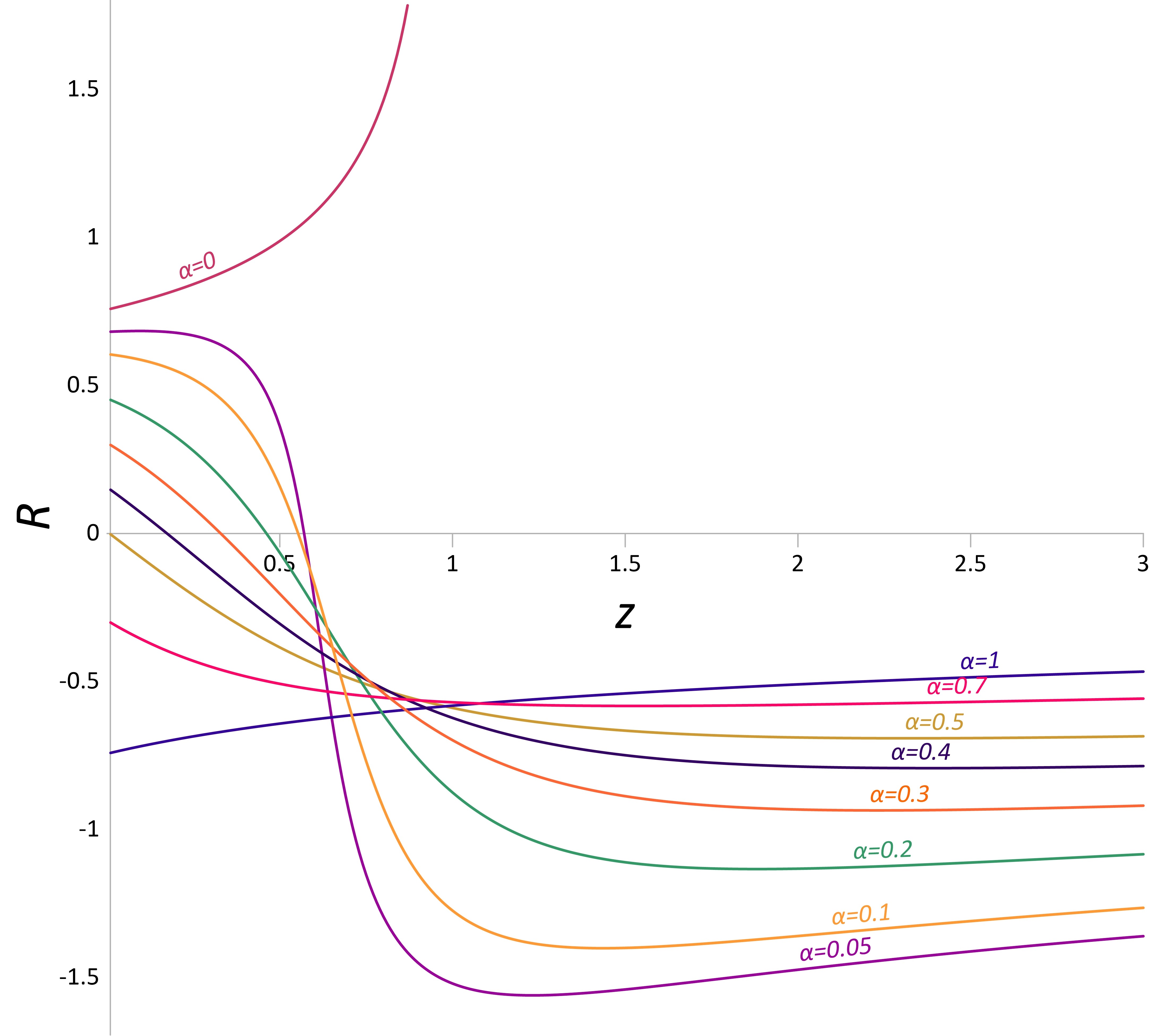}
    \caption{Thermodynamic curvature as a function of fugacity for isothermal processes (\(\beta = 1\)) in a 2-dimensional ideal \(\alpha\)-statistics gas with \(\sigma = 1\) (ultra-relativistic regime).}
    \label{fig3}
\end{figure}

For the ultra-relativistic regime in two dimensions, corresponding to \(D/\sigma = 2/1\), the thermodynamic curvature is plotted as a function of fugacity in Fig.~\ref{fig3}. It is evident that the same arguments discussed in the previous case (Fig.~\ref{fig2}) remain valid in this case as well.



Using these data, the plot of \( z^* \) as a function of \( \alpha \) is estimated for  non-relativistic particles.
\begin{figure}[h]
    \centering
    \includegraphics[scale=0.35]{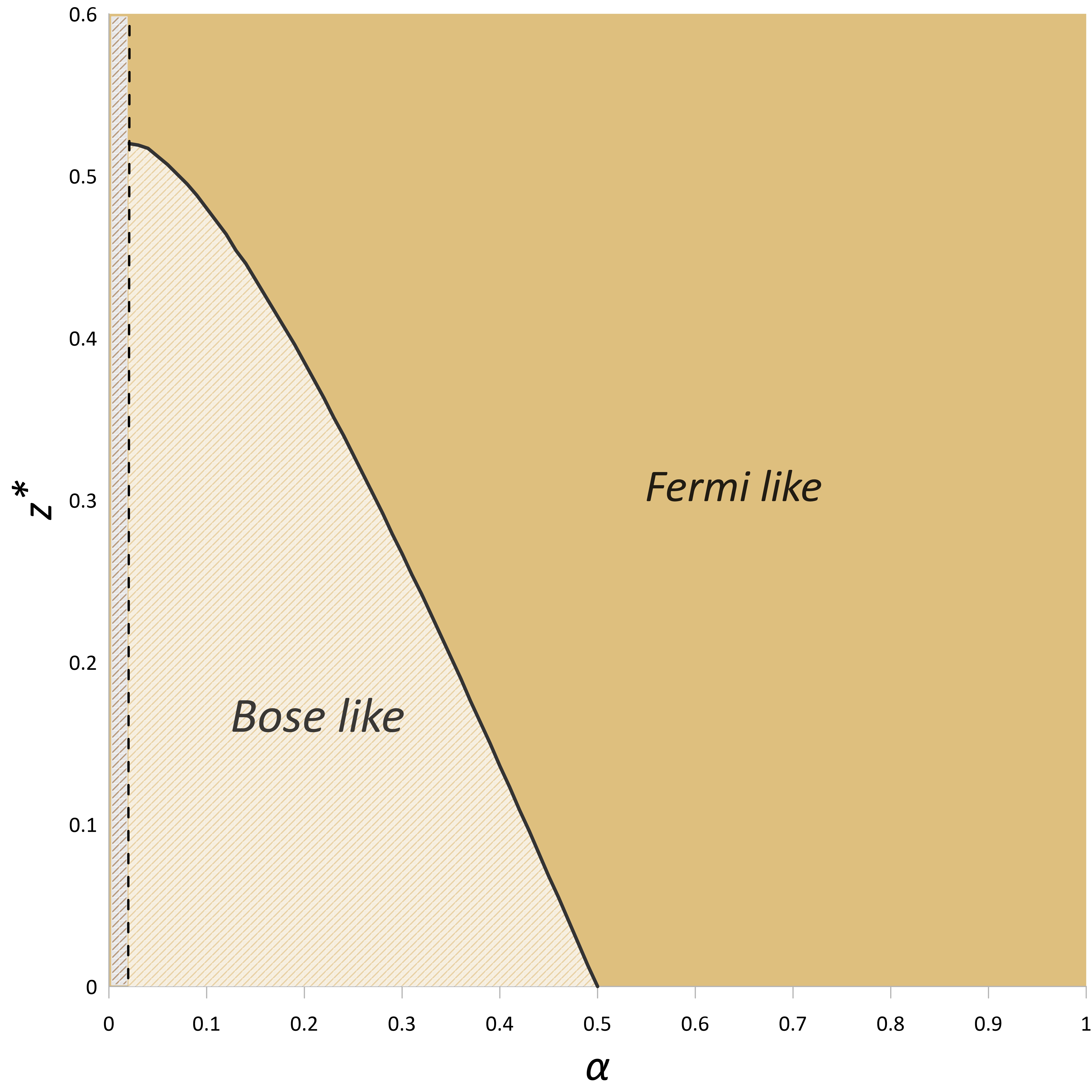}    
   \caption{The dependence of $z^*$ on $\alpha$ delineates two regions: a fermion-like region and a boson-like region. These regions are separated by the $z^*$ curve, corresponding to the points where the thermodynamic curvature changes sign.}
        \label{fig4}
\end{figure}
According to Figure (\ref{fig4}), in all dimensions, the thermodynamic curvature is negative (Fermi-like) for \( \alpha > 0.5 \) in all physical ranges. For \( \alpha < 0.5 \) the thermodynamic curvature is negative (Fermi-like) for the values \( z>z^{*} \) while it is positive for \( z<z^{*} \).In fact, the sign of thermodynamic curvature changes at $z=z^*$. For values of $\alpha < 0.02$, the accuracy of numerical computations deteriorates significantly. However, it is expected that the system behaves like bosons when $\alpha = 0$. Therefore, there is no specific value of $z$ at which the thermodynamic curvature changes sign. On the other hand, for a bosonic system, the value of $z$ is limited to its maximum value, namely $z_{\mathrm{max}} = 1$. Hence, in the region not plotted in Figure~(\ref{fig4}) due to the low accuracy of numerical calculations, it is expected that the curve of $z^{*}$ increases smoothly up to the maximum value of $z$.

\section{Crossover TEMPERATURES}\label{7}
Now, we obtain the value of \( z^* \) for each value of \( \alpha \). For a fixed number of particles, the corresponding temperature \( T^* \) can be determined using Eq.~(\ref{ne}). Thus, we find:
\begin{equation}
k_B T^* = \frac{h^2}{2m\pi} \left[ \frac{N}{V f_{\alpha}(z^*)} \right]^{\frac{2}{3}},
\end{equation}
\begin{widetext}
where the function \( f_{\alpha}(z) \) is defined as:
\begin{equation}
f_{\alpha}(z) = \frac{1}{\Gamma\left(\frac{D}{\sigma}\right)}
\int_{w_0}^{\infty} \frac{1}{w} \left[ \ln\left( (w + 1)^{1-\alpha} (w - 1)^{\alpha} z \right) \right]^{\frac{D}{\sigma} - 1}
\frac{w + 2\alpha - 1}{w^2 - 1} \, dw.
\end{equation}
\end{widetext}
The thermodynamic curvature changes sign at \( T = T^* \). To express this temperature in a dimensionless form, we introduce the ratio \( T / T_c \), where \( T_c \) is the well-known Bose–Einstein condensation temperature~\cite{pathria2011statistical}, defined as:
\begin{equation}
k_B T_c = \frac{h^2}{2m\pi} \left[ \frac{N}{V \zeta(3/2)} \right]^{\frac{2}{3}},
\end{equation}
where \( \zeta(3/2) \) is the Riemann zeta function. Therefore, we obtain:
\begin{equation}
\frac{T^*}{T_c} = \left[ \frac{\zeta(3/2)}{f_\alpha(z^*)} \right]^{\frac{2}{3}}.
\end{equation}

\begin{figure}[h]
    \centering
    \includegraphics[scale=0.35]{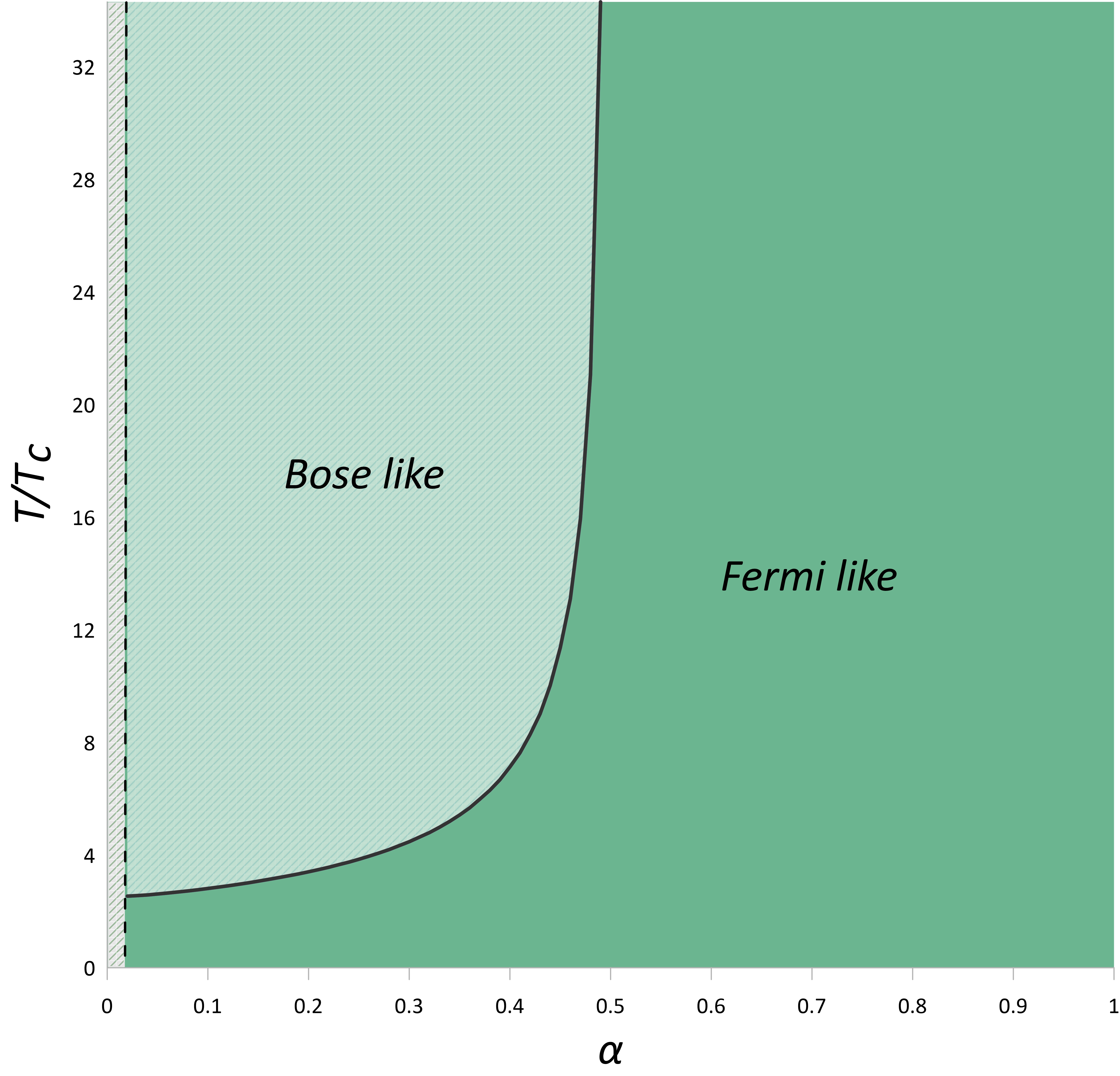}    \caption{Two distinct regions at different temperatures are separated by the sign change of the thermodynamic curvature. The sign of the thermodynamic curvature characterizes the nature of intrinsic statistical interactions: light and dark regions correspond to attractive (boson-like) and repulsive (fermion-like) interactions, respectively. The line represents the temperature at which the thermodynamic curvature changes sign, marking a crossover from negative to positive values.}
    \label{fig5}
\end{figure}

Figure (\ref{fig5}) illustrates the set of points where the thermodynamic curvature changes sign. The light region denotes a domain of positive curvature, associated with attractive intrinsic statistical interactions, while the darker region corresponds to negative curvature, indicating repulsive interactions.
For values of $\alpha < 0.02$, the accuracy of numerical computations deteriorates significantly. Nevertheless, the system is expected to exhibit bosonic behavior in the limit $\alpha \to 0$. Consequently, there should be no specific temperature at which the thermodynamic curvature changes sign. In the region not shown in Figure~\ref{fig5}, due to the limited accuracy of the numerical results, it is expected that the $T^{*}/T_c$ curve decreases smoothly to its minimum value at $T^{*}/T_c = 1$, corresponding to the normal phase of an ideal Bose gas. It is known that for $\alpha = 0$ and $T/T_c \leq 1$, the system enters the condensate phase. In the condensate phase, the thermodynamic curvature vanishes due to the fixed value of the fugacity, which reduces the thermodynamic parameter space to a trivially one-dimensional flat manifold.

\section{Negative Alphas}\label{8}
\begin{figure}[h]
    \centering
    \includegraphics[scale=0.35]{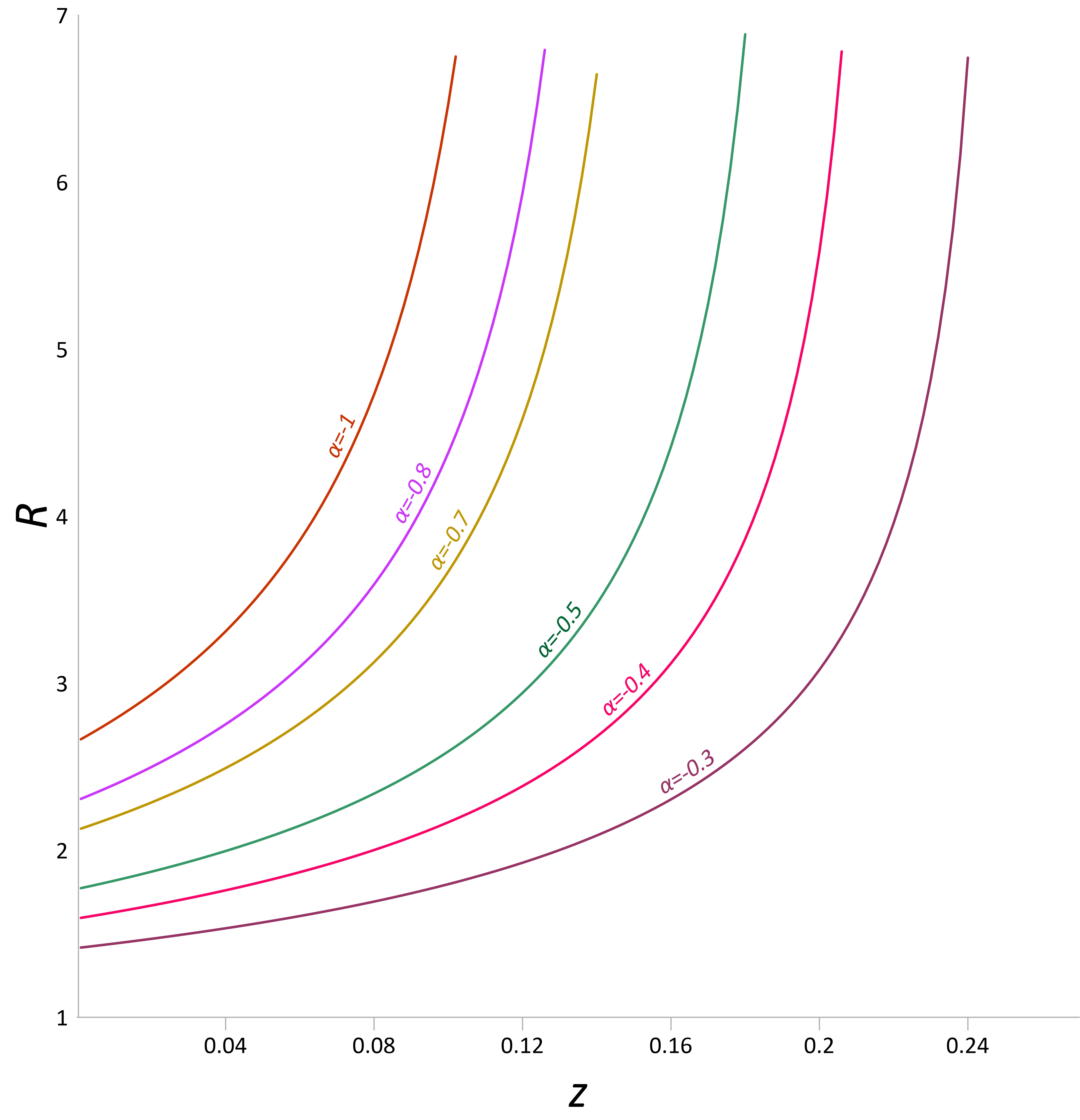}
    \caption{Thermodynamic curvature for some negative values of $\alpha$ as a function of fugacity for isothermal processes ($\beta=1$).}
     \label{Negetive}
\end{figure}

The regime of negative $\alpha$ values is of particular interest. Given the proposed relation for generalizing the state occupation weight function, it was evident that the value $\alpha = 0$ corresponds to bosons and yielded the maximum weight when considering only positive $\alpha$ values. If negative $\alpha$ values are also permitted, the weight function can attain values greater than even those in the bosonic case. We compute and analyze the thermodynamic curvature for certain negative $\alpha$ values. Figure \ref{Negetive} shows that the thermodynamic curvature as a function of fugacity is singular at specified values of fugacity for negative values of $\alpha$.

Previous studies established that Bose-Einstein condensation (BEC) in standard bosonic systems occurs at critical fugacity $z_c = 1$ \cite{janyszek1990riemannian}. Moreover, it has been demonstrated that a condensation phenomenon arising from generalized statistical distributions occurs precisely at the singularity of the thermodynamic curvature \cite{mirza2011condensation,mirza2010thermodynamic,mirza2011thermodynamic,esmaili2024thermodynamic,seifi2025intrinsic}. However, our analysis reveals that for $\alpha < 0$, condensation emerges at substantially reduced critical fugacities ($z_c < 1$), indicating fundamental alterations in phase transition criteria. This deviation suggests a modified density of states or emergent effective interactions inherent to hyperbosonic statistics. Crucially, while thermodynamic curvature in ideal Bose gases exhibits singular behavior exclusively at $z=1$ (corresponding to BEC criticality), systems with $\alpha < 0$ display curvature singularities at specific $z<1$ values. No analogous singularities occur for $\alpha > 0$. These observed singular points provide compelling evidence for phase transitions analogous to BEC at critical fugacity $z=z_{c}$ . Future investigations will comprehensively examine the statistical mechanics of negative $\alpha$ regimes, including impacts on heat capacity, distribution functions, and quantum effects, while further analysis of thermodynamic curvature will clarify the nature of effective interactions and phase transition mechanisms in these unconventional systems.
In our next work, we will conduct a more detailed investigation into the statistical behavior for negative \(\alpha\) and its impact on various thermodynamic properties, such as heat capacity, distribution function, and quantum effects. A comprehensive analysis of thermodynamic curvature in these systems could further clarify the nature of effective interactions and phase transitions under the influence of a negative \(\alpha\).

\section{Conclusion}\label{9}
In this work, we have introduced and investigated a generalized quantum statistical model termed \(\alpha\)-statistics, which smoothly interpolates between Bose--Einstein and Fermi--Dirac statistics and allows for a broader class of statistical behaviors including hyperbosonic regimes with \(\alpha < 0\). The formulation is constructed by modifying the occupation weight function, with the parameter \(\alpha\) governing the statistical interpolation. This approach is conceptually connected to Haldane's exclusion statistics, providing a geometric and thermodynamic realization of generalized exclusion principles through the curvature of the thermodynamic parameter space.

The central focus of our analysis has been the thermodynamic curvature, a geometric quantity sensitive to the nature of intrinsic statistical interactions. We showed that the sign of the curvature distinguishes between bosonic-like (attractive) and fermionic-like (repulsive) behavior, and that it changes at a characteristic crossover temperature \(T^*\). This temperature was expressed relative to the Bose--Einstein condensation temperature \(T_c\), and its dependence on \(\alpha\) was computed. Our results indicate that the ratio \(T^*/T_c\) decreases as \(\alpha\) approaches zero from above, consistent with the known behavior of ideal bosons entering the condensate phase below \(T_c\). In contrast to conventional cases, the thermodynamic curvature vanishes identically in the condensate phase due to the fixed fugacity, leading to a flat thermodynamic geometry.

A particularly intriguing aspect of \(\alpha\)-statistics arises in the negative-\(\alpha\) domain. In this regime, which corresponds to an enhancement of bosonic behavior beyond that of standard bosons, we observed singularities in the thermodynamic curvature at specific fugacities \(z_c < 1\), indicating the onset of condensation-like phase transitions distinct from the standard Bose--Einstein condensation at \(z = 1\). These singularities point to a modified density of states and possibly emergent effective interactions intrinsic to the statistical framework, signaling a new class of critical phenomena governed by hyperbosonic statistics.

The exploration of negative values of \(\alpha\) is currently ongoing and is expected to shed light on the nature of quantum correlations, critical behavior, and nontrivial phase structures in systems beyond the conventional fermion-boson dichotomy. Further work will address thermodynamic quantities such as heat capacity, particle distribution functions, and quantum coherence properties under \(\alpha < 0\). In particular, identifying universal features of phase transitions in these regimes could offer new insights into the role of generalized statistics in strongly correlated systems, low-dimensional quantum gases, or topologically ordered phases.

Overall, the \(\alpha\)-statistical framework presents a flexible and conceptually rich extension of quantum statistical mechanics. By unifying features of bosonic, fermionic, and exotic statistics within a single parameterized model, it opens new theoretical avenues for understanding emergent quantum behaviors in interacting many-body systems. Moreover, the geometric approach through thermodynamic curvature offers a powerful diagnostic tool for characterizing statistical interactions and criticality, and may have broader applications in quantum thermodynamics, information geometry, and condensed matter theory.

\bibliography{refs}
\end{document}